\documentclass[onecolumn,preprintnumbers,amsmath,aps]{revtex4-2}
\usepackage{graphicx}
\usepackage{dcolumn}
\usepackage{bm}
\usepackage{subfigure}
\usepackage{color}
\usepackage{microtype}
\usepackage{amsmath}
\usepackage{mathtools}
\usepackage{amssymb}
\begin{document}
%%%%%%%%%%%%%%%%%%%%%%%%%%%%

%%%%%%%%%%%%%%%%%%%%%%%%%%%%
\title{Cosmology in the mimetic higher-curvature $f(R,R_{\mu\nu}R^{\mu\nu})$ gravity}
%%%%%%%%%%%%%%%%%%%%%%%%%%%%
\author{Adam Z. Kaczmarek$^{1}$}\email{adamzenonkaczmarek@gmail.com}
\author{Dominik Szcz{\c{e}}{\'s}niak$^{2}$}
%%%%%%%%%%%%%%%%%%%%%%%%%%%%
\affiliation{$^1$Department of Physics, Cz{\c{e}}stochowa University of Technology, 19 Armii Krajowej Ave., 42200 Cz{\c{e}}stochowa, Poland}
\affiliation{$^2$ Department of Theoretical Physics, Faculty of Science and Technology, Jan D{\l}ugosz University in Cz{\c{e}}stochowa, 13/15 Armii Krajowej Ave., 42200 Cz{\c{e}}stochowa, Poland}
%%%%%%%%%%%%%%%%%%%%%%%%%%%%
\date{\today} 
\begin{abstract}
%%%%%%%%%%%%%%%%%%%%%%%%%%%%%%%%%%%%%%%%%%%%%%%%%%%%%%%%%%%%%%
In the framework of the mimetic approach, we study the $f(R,R_{\mu\nu}R^{\mu\nu})$ gravity with the Lagrange multiplier constraint and the scalar potential. We introduce field equations for the discussed theory and overview their properties. By using the general reconstruction scheme we obtain the power law cosmology model for the $f(R,R_{\mu\nu}R^{\mu\nu})=R+d(R_{\mu\nu}R^{\mu\nu})^p$ case as well as the model that describes symmetric bounce. Moreover, we reconstruct model, unifying both matter dominated and accelerated phases, where ordinary matter is neglected. Using inverted reconstruction scheme we recover specific $f(R,R_{\mu\nu}R^{\mu\nu})$ function which give rise to the de-Sitter evolution. Finally, by employing the perfect fluid approach, we demonstrate that this model can realize inflation consistent with the bounds coming from the BICEP2/Keck array and the Planck data. Thus, it is suggested that the introduced extension of the mimetic regime may describe any given cosmological model.
%%%%%%%%%%%%%%%%%%%%%%%%%%%%%%%%%%%%%%%%%%%%%%%%%%%%%%%%%%%%%%
\end{abstract}

\maketitle

%%%%%%%%%%%%%%%%%%%%%%%%%%%%%%%%%%%%%%%%%%%%%%%%%%%%%%%%%%%%%%
\section{Introduction}
%%%%%%%%%%%%%%%%%%%%%%%%%%%%%%%%%%%%%%%%%%%%%%%%%%%%%%%%%%%%%%

In the 1990s, the field of cosmology encountered serious breakthrough with the discovery of accelerated expansion of the Universe. This led to the current evolutionary status that is determined by the three contributions to the total density parameter ($\Omega$): the ordinary matter ($\Omega_m\approx5\%$), the dark matter ($\Omega_{DM}\approx 27\%$) and the dark energy ($\Omega_{DE} \approx 68\%$), where the latter contribution is responsible for the accelerated expansion phenomena \cite{riess1998}. Importantly, the nature of the {\it dark sector} in the energetic budget is also the reason for intense experimental and theoretical studies that exposed shortcomings of the general relativity (GR) \cite{huterer2017,copeland2006}. In this respect, one of the most promising proposals to deal with the mentioned issues is based on the extensions and modifications of the Einstein's framework of GR \cite{joyce2016}. Note that the proposals of this type are quite successful not only in the context of the aforementioned cosmic expansion but also in other issues of the modern cosmology \cite{nojrii2017}. Moreover, in terms of the quantum gravity, the standard GR cannot be renormalized but higher-curvature terms in the gravitational action may tackle this issue \cite{Julve1978,kiefer2006,Hamber2009}.  

In the literature the GR has been extended in many ways, mainly by including additional curvature terms, scalar fields or coupling with matter. For a well-written reviews on the modified gravities we refer reader to \cite{capozziello2011,nojiri2011,clifton2012,joyce2016}. Among many interesting approaches, one of the most popular proposals is the so-called $f(R)$ gravity introduced by Buchdachl in 70's \cite{buchdachl1970,sotiriou2010,bohmer2008}, where standard GR action is generalized to the arbitrary function of the Ricci scalar ($R\rightarrow f(R)$). Further curvature invariants were included in the $f(\mathcal{G})$ gravity, where arbitrary function of the Gauss-Bonnet invariant is added to the Lagrangian \cite{nojiri2005}. In a wide class of the higher-curvature models, the $f(R,\mathcal{G})$, $f(R,R_{\mu\nu}R^{\mu\nu})$ or theories with the higher-order of derivatives have also been introduced and studied extensively \cite{elizalde2010,de_la_Cruz_Dombriz2012,sharif2016,lambiase2015,easson2004,sharif2016}. The authors of \cite{caroll2005} considered combinations of the Ricci scalar, the Ricci squared ($R_{\mu\nu}R^{\mu\nu}$) and the Riemann squared ($R_{\mu\nu\alpha\beta}R^{\mu\nu\alpha\beta}$) terms and shown that such models can provide alternative to the dark energy. Moreover, theories of this type were recently studied in the context of divergences and renormalization group frameworks and play important role in the perturbative approach to the quantum gravity \cite{Avramidy1985,Hamber2009,Ohta2018}. Note that the general relativity can be further extended by the scalar-geometry or scalar-matter couplings. In this case gravitational action includes terms with contributions not only from geometry but also various functions of the scalar field, matter Lagrangian or trace of the energy-momentum tensor. Representative example of the scalar-geometry coupling is the Jordan-Brans-Dicke theory \cite{jordan1955,brans1961}, further generalized to the $f(\phi,R)$ and $f(\phi,\mathcal{G})$ gravity \cite{capozziello2011}. For the matter-geometry couplings, the $f(R,T)$ and $f(R,T,R_{\mu\nu}T^{\mu\nu})$ models ($T$ denotes trace of the energy-momentum tensor $T_{\mu\nu}$) with further extensions are good examples studied widely in the literature \cite{haghani2013,odintsov2013,harko2011,houndjo2012,harko2014b,Sharif2013,Kaczmarek2020}.

On the other hand, the dark matter is even more mysterious since very few information is known except the fact that it modifies galactic rotation, it is not baryonic, has no electric and color charges and do not interact with the light \cite{oikonomou2007,shafi2015}. Apart from the particle candidates for the dark matter, such as supersymmetry neutralino, an interesting candidate has been provided by Chamseddine and Mukhanov. In their so-called mimetic modification of the GR, the leading idea was to isolate gravity's conformal degree of freedom by parametrizing physical metric ($g_{\mu\nu}$) in terms of the scalar field ($\phi$) and the auxiliary metric ($\hat{g}_{\mu\nu}$) \cite{Chamseddine2013}. As a consequence, the conformal degree of freedom turns out to be dynamical, even in the absence of the standard matter fluids, and may mimic behavior of the cold dark matter. It is worth to remark here, that the mimetic theory has been recently extensively studied in various contexts \cite{dutta2018,gorji2020,golovnev2014,maleb2015,ganz2019,Bezerra2019,chamseddine2019,Mansoori2021,Cardenas2021}. Moreover, the mimetic approach has been combined with other modifications of gravity, such as $f(R)$, $f(\mathcal{G})$ or $f(R,T)$ theories with emphasis on the cosmological applications \cite{odintsov2015a,oikonomou2015,zhong2018,leon2015,haghani2018,gorji2018,baffou2017,Bhattacharjee2020}. The unimodular extensions of the mimetic approach were also introduced and studied in recent articles \cite{Odintsov2016,nojiri2016}. For introduction and review of the mimetic theory see \cite{sebastiani2017}.  

Besides the riddle of the dark Universe, the modified gravity is also important in terms of the early stages of the Universe. In this context, the theory of inflation is by now most possible scenario that describes early Universe history \cite{Linde2007,lyth1999}. This framework, developed in the 80's, provides explanation of important issues in the Big Bang cosmology such as the horizon, flatness and monopole problems \cite{guth1981}. According to the inflation paradigms, early stages of the Universe exhibited exponentially accelerated expansion, which after deceleration led to the standard cosmological eras \cite{hobson2006}. However, no direct proof that this early epoch occurred is given up to date. One should note that inflation is well described by the {\it slow-roll} models which usually contain single scalar field (inflaton) that drives inflation. Moreover, the form of the scalar field potential can be related to the spectral index as well as to the scalar and tensor perturbations as generated during the inflationary epoch. Thus, scalar potential may be consistent with observations. Recent observational data coming from Planck and BICEP2/Keck array heavily constrained scalar-to-tensor ratio and spectral index of primordial curvature perturbations corresponding to the inflation \cite{ade2016}. One should note, that the bounce paradigm may also be viable characteristic of the early Universe and an interesting alternative   \cite{bars2014,brandenberger1993,novello2008,cai2014}. From the modified gravity perspective, inflation can be successfully described and the unification of early-time and late-time acceleration eras is also possible \cite{nojiri2011}. In the literature, inflation in the modified gravity is still heavily studied and debated topic. In particular, for the $f(R)$ gravities a successful model of the Starobinsky's inflation exists, which is consistent with the local and large scale constraints of gravity \cite{starobinsky1980}. Alternatively, other modifications of the GR were studied in the context of inflation as well as the unification of accelerated phases of the Universe \cite{nojrii2017,cognola2008}. Note that inflation for the mimetic gravity with its extensions, was also recently investigated \cite{zhong2018,nojiri2014,odintsov2018,lambiase2015,Mansoori2021}.

In the context of the above, herein we combine mimetic idea with the general higher-curvature $f(R,R_{\mu\nu}R^{\mu\nu})$ gravity. We intent to extend previously considered mimetic framework with more general $f(R,R_{\mu\nu}R^{\mu\nu})$ gravity, with respect to the cosmological reconstruction of various models. After brief discussion of the field equations, by using the reconstruction method, we obtain the power law evolution model, the symmetric bounce as well as the transition between radiation and matter dominated phases. Then, in reference the current experimental data, we discuss inflationary model of the mimetic $f(R,R_{\mu\nu}R^{\mu\nu})$ gravity. By using the reconstruction methods, we obtain Lagrange multiplier and potential responsible for inflationary evolution described by the scale factor $H(N)=\alpha e^{\beta N}+\gamma $. Finally by employing the perfect fluid approach, we show that it is possible to achieve inflationary evolution in the discussed theory, in agreement with the current data.

%%%%%%%%%%%%%%%%%%%%%%%%%%%%%%%%%%%%%%%%%%%%%%%%%%%%%%%%%%%%%%
\section{Theoretical Model}
%%%%%%%%%%%%%%%%%%%%%%%%%%%%%%%%%%%%%%%%%%%%%%%%%%%%%%%%%%%%%%

The main idea behind the mimetic gravity is that the $g_{\mu\nu}$ metric (the fundamental variable in gravity) may be expressed by the new degrees of freedom \cite{lambiase2015}. By doing so, such degrees of freedom can admit wider class of solutions than the standard GR. In particular, the conformal degree of freedom can be isolated by expressing the physical metric in terms of the scalar field $\phi$ and the auxiliary metric $\hat{g}_{\mu\nu}$ as follows:
%%%%%%%%%%
\begin{align}
    g_{\mu\nu}=-\hat{g}^{\alpha\beta}\partial_{\alpha}\phi\partial_{\beta}\phi \hat{g}_{\mu\nu}.
\label{eq:refname1}    
\end{align}
Note that the above parametrization posses Weyl symmetry under the conformal transformation, {\it i.e.} $\hat{g}_{\alpha\beta}=e^{\omega(x)}g_{\alpha\beta}$. It is also worth to remark, that instead of using the physical metric $g_{\mu\nu}$ in variation of the gravitational action, one can use variation with respect to the metric $\hat{g}_{\mu\nu}$ and the scalar field $\phi$. As a result, from Eq. (\ref{eq:refname1}) the following constraint can be obtained:
\begin{align}
    g^{\mu\nu}(\phi,\hat{g}_{\mu\nu})\partial_{\mu}\phi\partial_{\nu}\phi=-1.
 \label{eq:refname2}    
\end{align}
%%%%%%%%%%
In order to impose given constraint at the action level we use the Lagrange multiplier formalism \cite{capoziello2010,makarenko2016}. In the mimetic regime we need to introduce the Lagrange multiplier ($\lambda$) that corresponds to the mimetic constraint \cite{lambiase2015}. Hence, for the mimetic $f(R,R^{\mu\nu}R_{ \mu\nu})$ gravity we have:
%%%%%%%%%%
\begin{align}
    S=\int d^4x\Big[\sqrt{-g}\big( f(R,R_{\mu\nu}R^{\mu\nu})-V(\phi)+\lambda(g^{\mu\nu}\partial_\mu\phi\partial_\nu\phi+1)\big) \Big],
\label{eq:refname3}
\end{align}
%%%%%%%%%%
where 
$f(R,R^{\mu\nu}R_{ \mu\nu})$ is the analytical function of the Ricci scalar $R$ and contraction of the Ricci tensors (the Ricci squared term) $R^{\mu\nu}R_{ \mu\nu}$. Moreover, $V(\phi)$ denotes scalar (mimetic) potential. We note that the equations obtained from the variation with respect to the physical metric $g_{\mu\nu}$ with imposed mimetic constraint are fully equivalent to the equations that one can derive by using action written in terms of the auxiliary metric $\hat{g}_{\mu\nu}$. We emphasize that standard matter fields are not part of our considerations since one of the core characteristics of the mimetic theory is to \textit{mimic} matter content \cite{lambiase2015}. In what follows, the variation of the action (\ref{eq:refname3}) with respect to the components of the metric tensor $g^{\mu\nu}$, gives the following field equation:
%%%%%%%%%%
\begin{align}\nonumber
f_R R_{\mu\nu}&-\nabla_{\mu}\nabla_{\nu}f_R+g_{\mu\nu}\Box f_R+\Box( R_{\mu\nu}f_Y)+g_{\mu\nu}\nabla^{\alpha}\nabla^{\beta}( R_{\alpha\beta}f_Y)-2\nabla_{\alpha}\nabla_{\beta}(R^{\alpha}_{(\mu}\delta^{\beta}_{\nu)}f_Y)+2R^{\alpha}_{\mu}R_{\alpha\nu}f_Y\\ 
&-\frac{1}{2}g_{\mu\nu}(f-V(\phi)+\lambda(g^{\alpha\beta}\partial_\alpha\phi\partial_\beta\phi+1))+\lambda \partial_\mu\phi\partial_\nu\phi=0,
\label{eq:refname4} 
\end{align}
%%%%%%%%%%
where $Y=R_{\mu\nu}R^{\mu\nu}$, $f_Y=\frac{\partial f(R,Y)}{\partial R}$ and $f_Y=\frac{\partial f(R,Y)}{\partial Y}$, whereas the indices in brackets denote symmetrization ($A_{(\mu\nu)}=\frac{1}{2}(A_{\mu\nu}+A_{\nu\mu})$). Next, by varying the action with respect to the scalar field $\phi$ the following scalar equation can be obtained:
%%%%%%%%%%
\begin{align}
        2\nabla^{\mu}(\lambda \partial_{\mu}\phi)+V'(\phi)=0,
\label{eq:refname5}         
\end{align}
%%%%%%%%%%
where prime stands for the differentiation with respect to the scalar field ($V'(\phi)=\frac{dV(\phi)}{d\phi}$). It is important to remark, that the variation of the action with respect to $\lambda$ returns mimetic constraint.

Further in this work we assume that the background geometry is described by the FLRW metric, with the following line element:
%%%%%%%%%%
\begin{align}
    ds^2=-dt^2+a^2(t)dx_idx^i,
\label{eq:refname6}
\end{align}
%%%%%%%%%%
where $i=1,2,3$ and $a(t)$ is the scale factor. For simplicity, we also assume that the scalar field depends only on time, {\it i.e.} $\phi=\phi(t)$. By using the FLRW metric (\ref{eq:refname6}), the corresponding $(t,t)$ component of the mimetic $f(R,R_{\mu\nu}R^{\mu\nu})$ gravity equations (\ref{eq:refname4}) is:
%%%%%%%%%%
\begin{align}\nonumber
    -3(\Dot{H}+H^2)f_R&+3 H \partial_t f_R+\frac{1}{2}(f+\lambda(\dot{\phi}^2+1)- V(\phi))+6(2 H \ddot{H}-2\dot{H}^2+3H^2\dot{H}-3H^4)f_Y\\ &+6(2 H \dot{H}+3H^3)\partial_t f_Y=0,  
    \label{eq:refname7}
\end{align}
%%%%%%%%%%
while the $(i,j)$ component reads:
%%%%%%%%%%
\begin{align} \nonumber
    &(\dot{H}+3H^2)f_R-2 H \partial_t f_R-\partial_{tt}f_R-\frac{1}{2}(f-\lambda(\dot{\phi}^2-1)-V(\phi))-2(2\dddot{H}+12 H \ddot{H}-9H^4+9H^2\dot{H}+6\dot{H}^2)f_Y
\\ 
&-2(18 H\dot{H}+6H^3+4\ddot{H})\partial_t f_Y-2(2\dot{H}+3H^2)\partial_{tt}f_Y=0.
\label{eq:refname8}
\end{align}
%%%%%%%%%%
From Eq. (\ref{eq:refname5}) the corresponding scalar equation is given as:
%%%%%%%%%%%%
\begin{align}
    6H\lambda \dot{\phi}+2(\dot{\lambda}\dot{\phi}+\lambda\Ddot{\phi})-V'(\phi)=0.
    \label{eq:refname9}
\end{align}
%%%%%%%%%%
Note, that {\it dot} denotes differentiation with respect to the cosmic time $t$ and $H=\dot{a}/a$ is the Hubble rate. For brevity, we note that the Ricci scalar and the Ricci tensor squared in the FLRW spacetime are:
%%%%%%%%%%
\begin{align}
R=6(\dot{H}+2H^2),\;\;\;\;
    Y=R_{\mu\nu} R^{\mu\nu}=12\big(\dot{H}^2+3H^2(\dot{H}+H^2)\big).
    \label{eq:refname10}
\end{align}
%%%%%%%%%%

Finally, the mimetic constraint for the metric given by Eq. (\ref{eq:refname6}) is:
%%%%%%%%%
\begin{align}
    \dot{\phi}^2=1,
     \label{eq:refname11}
\end{align}
%%%%%%%%%
and yields:
%%%%%%%%%
\begin{align}
    \phi=t+C,
    \label{eq:refname12}
\end{align}
%%%%%%%%%%
with $C$ being integration constant. Keeping in mind that the scalar field appears only with derivatives, shift symmetry allows for setting constant $C=0$ without losing generality. By using $\phi=t$ the field and scalar equations (\ref{eq:refname7}-\ref{eq:refname9}) reduce to:
%%%%%%%%%%
\begin{align}\nonumber
    -3(\Dot{H}+H^2)f_R&+3 H \partial_t f_R+\frac{1}{2}(f- V(\phi))+6(2 H \ddot{H}-2\dot{H}^2+3H^2\dot{H}-3H^4)f_Y\\ &+6(2 H \ddot{H}+3H^3)\partial_t f_Y+\lambda=0,
    \label{eq:refname13}
\end{align}
%%%%%%%%%%
\begin{align} \nonumber
    &(\dot{H}+3H^2)f_R-2 H \partial_t f_R-\partial_{tt}f_R-\frac{1}{2}(f-V(\phi))-2(2\dddot{H}+12 H \ddot{H}-9H^4+9H^2\dot{H}+6\dot{H}^2)f_Y
    \\ 
    &-2(18H \dot{H}+6H^3+4\ddot{H})\partial_t f_Y-2(2\dot{H}+3H^2)\partial_{tt}f_Y=0,
    \label{eq:refname14}
\end{align}
and
%%%%%%%%%%
 \begin{align}
    6H\lambda +2\dot{\lambda}-V'(\phi)=0.
    \label{eq:refname15}
\end{align}
%%%%%%%%%%%
Hence, Eq. (\ref{eq:refname14}) can be used to determine potential $V(\phi)$ that generates evolution of the Hubble parameter for the specific $f(R,Y)$ gravity:
%%%%%%%%%%
\begin{align}\nonumber
   V(\phi=t)=&-2(\dot{H}+3H^2)f_R+4 H \partial_t f_R+2\partial_{tt}f_R+f+4\big[(2\dddot{H}+12 H \ddot{H}-9H^4+9H^2\dot{H}+8\dot{H}^2)f_Y
    \\ 
    &+(18H \dot{H}+6H^3+4\ddot{H})\partial_t f_Y+(2\dot{H}+3H^2)\partial_{tt}f_Y\big].
    \label{eq:refname16}
\end{align}
%%%%%%%%%%%
By combining Eqs. (\ref{eq:refname13}) and (\ref{eq:refname14}), the analytic form of the Lagrange multiplier $\lambda(t)$ is provided:
%%%%%%%%%%%
\begin{align}
    \lambda(t)=2\dot{H}f_R-H\partial_t f_R+ \partial_{tt}f_R +4(3 H \ddot{H}+\dddot{H}+\dot{H}^2)f_Y- 2(3H^3 -12 H \dot{H}-4 \ddot{H})\partial_t f_Y+2(2\dot{H}+3H^2)\partial_{tt}f_{Y}.
    \label{eq:refname17}   
\end{align}
%%%%%%%%%%%
This equation allows to obtain the desired evolution in various ways. Given the Hubble parameter $H(t)$ for the specified $f(R,Y)$ function, one can find corresponding multiplier $\lambda$ and mimetic potential $V(\phi)$. Also, given the form of the mimetic potential and the Lagrange multiplier $\lambda $, one can solve above equations to find proper $f(R,R_{\mu\nu}R^{\mu\nu})$ gravity model that realizes cosmological scenario of interest.

%%%%%%%%%%%%%%%%%%%%%%%%%%%%%%%%%%%%%%%%%%%%%%%%%%%%%%%%%%%%%%
\section{Reconstruction of the FLRW cosmologies}
%%%%%%%%%%%%%%%%%%%%%%%%%%%%%%%%%%%%%%%%%%%%%%%%%%%%%%%%%%%%%%

In the following section, we use reconstruction methods in order to obtain various cosmological scenarios in the mimetic $f(R,R_{\mu\nu}R^{\mu\nu})$ gravity. The procedure of cosmological reconstruction is widely used in the context of modified theories of gravity and helps with the extraction of physical implications in the theory of interest \cite{nojiri2014,nojiri2006,zubair2016,bamba2012,Hossienkhani2014,rudra2016,Carloni2012,Kaczmarek2020}.  In general, obtaining solutions for the GR's extensions may be troublesome, since even small modifications can drastically increase difficulty of equations. The reconstruction scheme used in the extensions of the GR relies on the fact that arbitrary function (or potential) is used in the definition of the modified gravity. Once the Hubble factor $H$ is specified, the field equations are solved to obtain the model of interest. This technique is a good way to survey modified theories of gravity, since satisfying well established cosmological models in gravity's extensions is desired. Moreover, this technique yields difference between the GR and its modifications, namely specific class (or classes) of the extended gravity may describe any given spacetime \cite{nojiri2011}. In this section, once the function $f(R,Y)$ is given, we obtain the Lagrange multiplier $\lambda$ and potential $V$ satisfying specified evolution. In general, by using this technique one can achieve model describing any given Hubble rate such as inflation or transitions between phases for the mimetic $f(R,R_{\mu\nu}R^{\mu\nu})$ approach.

In the present study, we assume specific form of the $f(R,R_{\mu\nu}R^{\mu\nu})$ function, where Ricci squared term enters with arbitrary power $p$, {\it i. e.}:
%%%%%%%%%%%
\begin{align}
  f(R,Y)=R + dY^p.
\label{eq:refname18}
\end{align}
%%%%%%%%%%%
Thus, the rearranged equations for this choice of the $f(R,Y)$ gravity are:
%%%%%%%%%%%
\begin{align}\nonumber
   &V(t)=-2(\dot{H}+3H^2)f_R+4 H \partial_t f_R+2\partial_{tt}f_R+f+4dp\big [(2\dddot{H}+12 H \ddot{H}-9H^4+9H^2\dot{H}+8\dot{H}^2)Y^{p-1}
    \\ 
    &+(18H \dot{H}+6H^3+4\ddot{H})(p-1) Y^{p-2}\dot{ Y} +(2\dot{H}+3H^2)((p-1)(p-2) Y^{p-3} \dot{Y}^2+(p-1) Y^{p-2} \ddot{Y})\big], 
    \label{eq:refname19}
\end{align}
%%%%%%%%%%%
and
%%%%%%%%%%%
\begin{align}\nonumber
    &\lambda(t)=2\dot{H}f_R-H\partial_t f_R+ \partial_{tt}f_R +4dp(3 H \ddot{H}+\dddot{H}+\dot{H}^2)Y^{p-1}- 2dp(3H^3 -12 H \dot{H}-4 \ddot{H})(p-1)Y
    ^{p-2}\dot{Y}\\ 
    &+2dp(2\dot{H}+3H^2)\big((p-1)(p-2) Y^{p-3} \dot{Y}^2+(p-1) Y^{p-2} \ddot{Y}\big).
    \label{eq:refname20}
\end{align}
%%%%%%%%%%%
As an example, let us consider the power-law scale factor and the corresponding Hubble parameter:
%%%%%%%%%%%
\begin{align}
    a(t)=a_0 t^{n},\;\;\; H(t)=\frac{n}{t}.
    \label{eq:refname21}
\end{align}
%%%%%%%%%%%
The solutions of this type are very useful in cosmology, especially in describing various phases in the history of the Universe \cite{hobson2006}. We note that for $0<n<1$, the decelerated universe occurs where for dust dominated Universe $n=\frac{2}{3}$ and for the radiation era $n=\frac{1}{2}$. The corresponding Ricci scalar and Ricci squared are:
%%%%%%%%%%%
\begin{align}
    R=-6\frac{n}{t^2}(1-2n),\;\;\;\;Y=12n^2\frac{(3n^2-3n+1)}{t^4}.
    \label{eq:refname22}
\end{align}
%%%%%%%%%%%
By using this expressions in the field equations, one can reconstruct the Lagrange multiplier and the mimetic potential for the $f(R,R_{\mu\nu}R^{\mu\nu}$ gravity. The general expressions obtained for the Lagrange multiplier and potential are given in the Appendix. For the dust $n=2/3$ and the inverse $p=-1$ power of $Y$, one gets:
%%%%%%%%%%%
\begin{align}
    \lambda(t)=d\frac{45}{8}t^4-\frac{4}{3t^2}
        \label{eq:refname23}
\end{align}
%%%%%%%%%%%
and
%%%%%%%%%%%
\begin{align}
    V(t)=d\frac{135}{16}t^4-\frac{8}{3t^2}.
        \label{eq:refname24}
\end{align}
%%%%%%%%%%%
It is worth to mention, that the Ricci squared contribution leads to the higher-order terms in the expressions for $V$ and $\lambda$. In the absence of this term ($d=0$), the potential takes form $\frac{8}{3}t^{-2}$ that is consistent with the result obtained in the standard mimetic gravity, {\it i.e.}  $V(t)=C_0t^{-2}$ for constant $C_0=8/3$ \cite{lambiase2015}.

As an another example we consider the exponential symmetric bounce cosmology, associated with the scale factor and the Hubble rate:
%%%%%%%%%%%
\begin{align}
    a(t)=e^{\alpha t^2},\;\;\;\; H(t)=2 \alpha t.
    \label{eq:refname25}
\end{align}
%%%%%%%%%%%
One of the main characteristics of the bounce cosmologies is the absence of the initial singularity. These theories constitute also interesting alternatives to the well-established inflationary paradigm \cite{Brandenberger2017,brandenberger1993}. For these scenarios, the Universe contraction decreases the effective radius of the Universe to the minimal size. Then, accelerated expansion occurs. In this case the Ricci scalar and Ricci tensor squared are:
%%%%%%%%%%%
\begin{align}
    R=-6 (2 \alpha + 8 t^2 \alpha^2),\;\;\;\; Y=48 \alpha^2 (1 + 6 t^2 \alpha + 12 t^4 \alpha^2).
    \label{eq:refname26}
\end{align}
%%%%%%%%%%%
The obtained Lagrange multiplier and mimetic potential take form:
%%%%%%%%%%%
\begin{align}
\lambda(t)=4\alpha    -\frac{d (3456 \alpha ^5
   t^{10}+11376 \alpha ^4
   t^8+8208 \alpha ^3 t^6+1788
   \alpha ^2 t^4+48 \alpha 
   t^2-11)}{144 \alpha
   ^2 (12 \alpha ^2 t^4+6
   \alpha  t^2+1)^4},
    \label{eq:refname27}
\end{align}
%%%%%%%%%%%
%%%%%%%%%%%
\begin{align}
  V(t)= -6 (2 \alpha + 8 t^2 \alpha^2) - 
  2 (2 \alpha + 12 t^2 \alpha^2)+    \frac{d (1152 \alpha ^6
   t^{12}+2976 \alpha ^5
   t^{10}+816 \alpha ^4
   t^8-472 \alpha ^3 t^6-124
   \alpha ^2 t^4+14 \alpha 
   t^2+3)}{16 \alpha ^2
   (12 \alpha ^2 t^4+6
   \alpha  t^2+1)^4},
    \label{eq:refname28}
\end{align}
%%%%%%%%%%%
when again we use $p=-1$ and the general form of the potential and the Lagrange multiplier is listed in the Appendix. 

In the following part, we provide reconstruction of the model that describes transition between matter-dominated and accelerated phases of the Universe's history. In this scenario, the Hubble rate is given by \cite{houndjo2012}:
%%%%%%%%%%%
\begin{align}
  H=g_0+\frac{g_1}{t}.
  \label{eq:refname29}
\end{align}
%%%%%%%%%%%
It is important to remark, that for small $t$, $H\approx \frac{g_1}{t}$ and the Universe is filled with the perfect fluid with the EoS (equation of state) parameter $w=-1+2/3g_1$. On the other hand, for the large $t$, the Hubble rate approaches $H\rightarrow g_0$ and the Universe looks like the de-Sitter one \cite{nojiri2011,houndjo2012}. We note that in this scenario, there is no real matter and the contribution of the mimetic $f(R,Y)$ gravity plays the role of the standard matter content.

In the small $t$ limit, the Lagrange multiplier and the mimetic potential are equal to:
%%%%%%%%%%%
%%%%%%%%%%%
\begin{align}
   \lambda(t)\approxeq\frac{12 d g_1{}^2 t^6-43 d
   g_1 t^6+30 d t^6-648
   g_1{}^8+1296 g_1{}^7-1080
   g_1{}^6+432 g_1{}^5-72
   g_1{}^4}{36 g_1{}^3 (3
   g_1{}^2-3 g_1+1){}^2
   t^2},
    \label{eq:refname30}
\end{align}
%%%%%%%%%%%
%%%%%%%%%%%
\begin{align}
  V(t)\approxeq  \frac{6 d g_1{}^3 t^6-19 d
   g_1{}^2 t^6-15 d g_1 t^6+20
   d t^6+648 g_1{}^9-1728
   g_1{}^8+1944 g_1{}^7-1152
   g_1{}^6+360 g_1{}^5-48
   g_1{}^4}{12 g_1{}^3 (3
   g_1{}^2-3 g_1+1){}^2
   t^2},
     \label{eq:refname31}
\end{align}
%%%%%%%%%%
while in the large $t$ limit one gets:
%%%%%%%%%%%
\begin{align}
\lambda(t)\approxeq0,\;\;\; V(t)\approxeq6g_0^2+\frac{d}{16g_0^4}
     \label{eq:refname32}
\end{align}
%%%%%%%%%%%
where again we have chosen $p=-1$. Note also that the full form of the obtained potential and the Lagrange multiplier can be found in the Appendix. Thus, in principle, the mimetic $f(R,R_{\mu\nu}R^{\mu\nu})$ gravity is able to unify phases dominated by matter with the transition to the late time accelerated evolution even in the absence of real matter for the appropriate mimetic potential and Lagrange multiplier. 

To this end, we reconstruct the particular class of $f(R,R_{\mu\nu}R^{\mu\nu})$ gravity that satisfy mimetic field equations, once particular mimetic potential and the Lagrange multiplier are assumed. This is inversion of the previously used reconstruction. In this approach, as an example, we consider the de-Sitter space as specified by the constant Hubble parameter:
%%%%%%%%%%%
\begin{align}
    H=H_0.
         \label{eq:refname33}
\end{align}
%%%%%%%%%%%
We remark that this spacetime usually serves as a description of the accelerated expansion phase and the early Universe inflation \cite{hobson2006}. By assuming the constant potential $V$ and the multiplier $\lambda$ as:
%%%%%%%%%%%
\begin{align}
    V(t)=\alpha H_0,\;\;\;\; \lambda(t)=2 \beta H_0,
    \label{eq:refname34}
\end{align}
%%%%%%%%%%%
we obtain the following partial differential equation (PDE):
%%%%%%%%%%%
\begin{align}
    3H_0^2 f_R+\frac{1}{2}f+6H_0^2+ \frac{(\beta-\alpha)}{2}H_0-18H_0^2f_Y=0.
    \label{eq:refname35}
\end{align}
%%%%%%%%%%%
This PDE has the following solution:
%%%%%%%%%%%
\begin{align}
    f(R,Y)=12H_0^2+(\alpha-\beta)H_0+e^{-\frac{R}{6H_0^2}} F(6H_0^2R+Y),
        \label{eq:refname36}
\end{align}
%%%%%%%%%%%
where $F$ is an arbitrary function.

In conclusion, the reconstruction technique can work in both ways: one can assume specific functional form of the mimetic $f(R,R_{\mu\nu}R^{\mu\nu})$ and cosmological evolution to obtain the mimetic potential and the Lagrange multiplier and \emph{vice versa}. In general, by using this procedure one can obtain any cosmological scenario for the mimetic $f(R,R_{\mu\nu}R^{\mu\nu})$ gravity.

%%%%%%%%%%%%%%%%%%%%%%%%%%%%%%%%%%%%%%%%%%%%%%%%%%%%%%%%%%%%%%
\section{Inflation}
%%%%%%%%%%%%%%%%%%%%%%%%%%%%%%%%%%%%%%%%%%%%%%%%%%%%%%%%%%%%%%

Inflation is believed to be one of the fundamental building blocks of the modern cosmology that may solve some of its problems \cite{guth1981,starobinsky1980}. This is to say, the main goal of the inflationary theory is to explain primordial fluctuations {\it i.e.} the density variations occurring in very early stages of the cosmic evolution. Since inflation is the most plausible scenario for the early Universe, the modified theories of gravity should be able to properly describe this phase of the cosmic history. Moreover, the deviations from GR can be interpreted as a quantum-induced corrections or motivated by the ultra-violet (UV) behavior of the quantum gravity, playing an important role in the inflationary phase of the early Universe \cite{bamba2015}. Thus, viable description of this stage in any modified gravity is desired. Therefore, in the following section we analyse inflation in the context of the mimetic $f(R,R_{\mu\nu}R^{\mu\nu})$ gravity to explore ability to obtain model comparable with the recent BICEP2/Keck observations \cite{ade2016,bamba2015}. Note, that the cosmological reconstruction can also be successfully applied to the discussed scenario. In this section we obtain the mimetic $f(R,R_{\mu\nu}R^{\mu\nu})$ gravity that describes inflationary model which is consistent with the recent observational data. Again, the functional form of Eq. (\ref{eq:refname18}) is assumed.

While discussing the inflationary cosmology it is worth to use the number of e-foldings $N$ (intervals for which the scale factor grows by the $N$ factors of $e$) instead of the cosmic time $t$. The relation between the scale factor and the e-folding number $N$ is following, $e^{N}=\frac{a}{a_0}$, where $a_0$ is the initial value of the scale factor in the initial time instance. For brevity, we list transformation rules for the time derivatives with respect to the e-foldings number:
%%%%%%%%%%%
\begin{align}\nonumber
    &\frac{d}{dt}=H(N) \frac{d}{dN},\;\;\;\; \frac{d^2}{dt^2}=H^2(N)\frac{d^2}{dN^2}+H(N)H'(N)\frac{d}{dN},\\
    &\frac{d^3}{dt^3}=3H^2(N)H'(N)\frac{d^2}{dN^2}+H^2(N)H''(N)\frac{d}{dN}+H(N)H'^2(N)\frac{d}{dN}+H^3(N)\frac{d^3}{dN^3}.
        \label{eq:refname37}
\end{align}
%%%%%%%%%%%
We note that prime symbols ($'$) correspond to the derivatives $\frac{dH(N)}{dN}$.

In order to study the {\it slow-roll} indices we use perfect fluid approach developed in \cite{bamba2014}. In this approach extra terms in the gravitational action (\ref{eq:refname1}) can be regarded as the perfect fluid. This formalism allows to obtain the spectral indices independently from the model. 
According to this approach, the {\it slow-roll} parameters are given as a functions of the Hubble rate $H$:
%%%%%%%%%%%
\begin{align}\nonumber
    \epsilon&=\frac{-4H(N)}{H'(N)}\Big(\frac{H'^2(N)+6H'(N)H(N)+H''(N)H(N)}{H'(N)H(N)+3H^2(N)}\Big)^2,\\ 
    \eta&=-\frac{\Big(9\frac{H'(N)}{H(N)}+3\frac{H''(N)}{H(N)}+\frac{1}{2}\Big(\frac{H'(N)}{H(N)}\Big)^2-\frac{1}{2}\Big(\frac{H''(N)}{H(N)}\Big)^2+\frac{H''(N)}{H'(N)}+\frac{H'''(N)}{H'(N)}\Big)}{2\big(3+\frac{H'(N)}{H(N)}\big)}.     
    \label{eq:refname38}
\end{align}
%%%%%%%%%%%
By using the above, the spectral index of the curvature perturbations $n_s$ and the scalar-to-tensor ratio $r$ in terms of the {\it slow-roll} parameters can be provided:
%%%%%%%%%%%
\begin{align}
    n_s=-6\epsilon+2\eta+1,\;\;\;\;r=16\epsilon.    \label{eq:refname39}
\end{align}

We remark that the spectral index describes variation of the density fluctuations with respect to the scale. On the other hand,  the scalar-tensor ratio relates spectras of the scalar and tensor perturbations. By using Eqs. (\ref{eq:refname16}), (\ref{eq:refname17}) and (\ref{eq:refname37}), the field equations in terms of the e-foldings number $N$ are:
%%%%%%%%%%%
\begin{align}\nonumber
    &\lambda(N)=H(N)^2 f_R''(N)-H(N)^3
   f_R'(N) H'(N)+H(N)
   f_R'(N) H'(N)+2
   f_R(N) H(N) H'(N)+4
   H(N)^3 f_Y''
   H'(N)\\ \nonumber
   &+12 H(N)^2
   f_Y'
   H'(N)^2+H(N)^3
   f_Y' \Big(8
   H''(N)+30 H'(N)\Big)-6
   H(N)^4
   \Big(f_Y'(N)-f_Y''(N)\Big)+4
   f_Y(N) H(N) H'(N)^3\\ 
   &+4
   f_Y H(N)^3
   \Big(H^{(3)}(N)+3
   H''(N)\Big)+16
   f_Y H(N)^2 H'(N)
   \Big(H''(N)+H'(N)\Big),
            \label{eq:refname40}
\end{align}
%%%%%%%%%%%
%%%%%%%%%%%
\begin{align}\nonumber
    &V(N)=f+2 H(N)^2 f_R''(N)+2
   H(N) f_R'(N) H'(N)+4
   H(N)^2 f_R'(N)-2
   f_R(N) H(N)
   \Big(H(N)
   (H''(N)+3)+H'(N)
   ^2\Big)\\ \nonumber
   &+8 H(N)^3
   f_Y''(N) H'(N)+12
   H(N)^4 f_Y''(N)+16
   H(N)^3 f_Y'(N)
   H''(N)+84 H(N)^3
   f_Y'(N) H'(N)+24
   H(N)^2 f_Y'(N) \\ \nonumber &\times
   H'(N)^2+24 H(N)^4
   f_Y'(N)+4
   f_Y(N) H(N) \Big(2
   H'(N)^3+2 H(N)^2
   (H^{(3)}(N)+6
   H''(N)\Big)+4 H(N) H'(N)
   \Big(2 H''(N)\\&+3
   H'(N)\Big) -9
   H(N)^3),
      \label{eq:refname41}
\end{align}
%%%%%%%%%%%
where we have used relations from Eqs. (\ref{eq:refname37}). Thus, one can obtain the Lagrange multiplier and potential corresponding to the scale factor $H(N)$, effectively reconstructing any given inflationary model. We note, that inverse reconstruction method is also possible once $\lambda(N)$ and $V(N)$ are specified.

\begin{figure}[!htb]
   \begin{minipage}{0.5\textwidth}
     \centering
     \includegraphics[width=1.
     \linewidth]{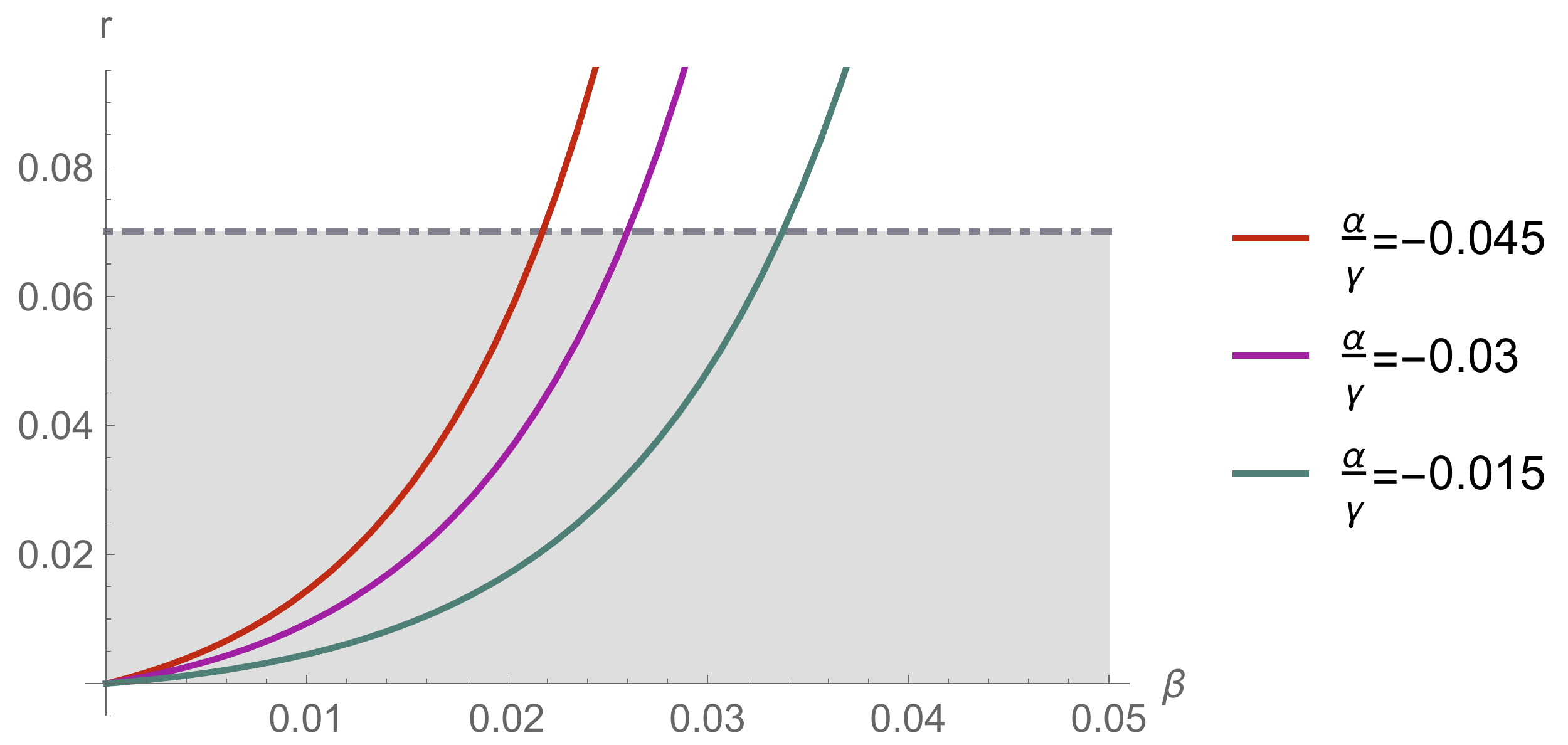}
   \end{minipage}\hfill
   \begin{minipage}{0.5\textwidth}
     \centering
     \includegraphics[width=1.\linewidth]{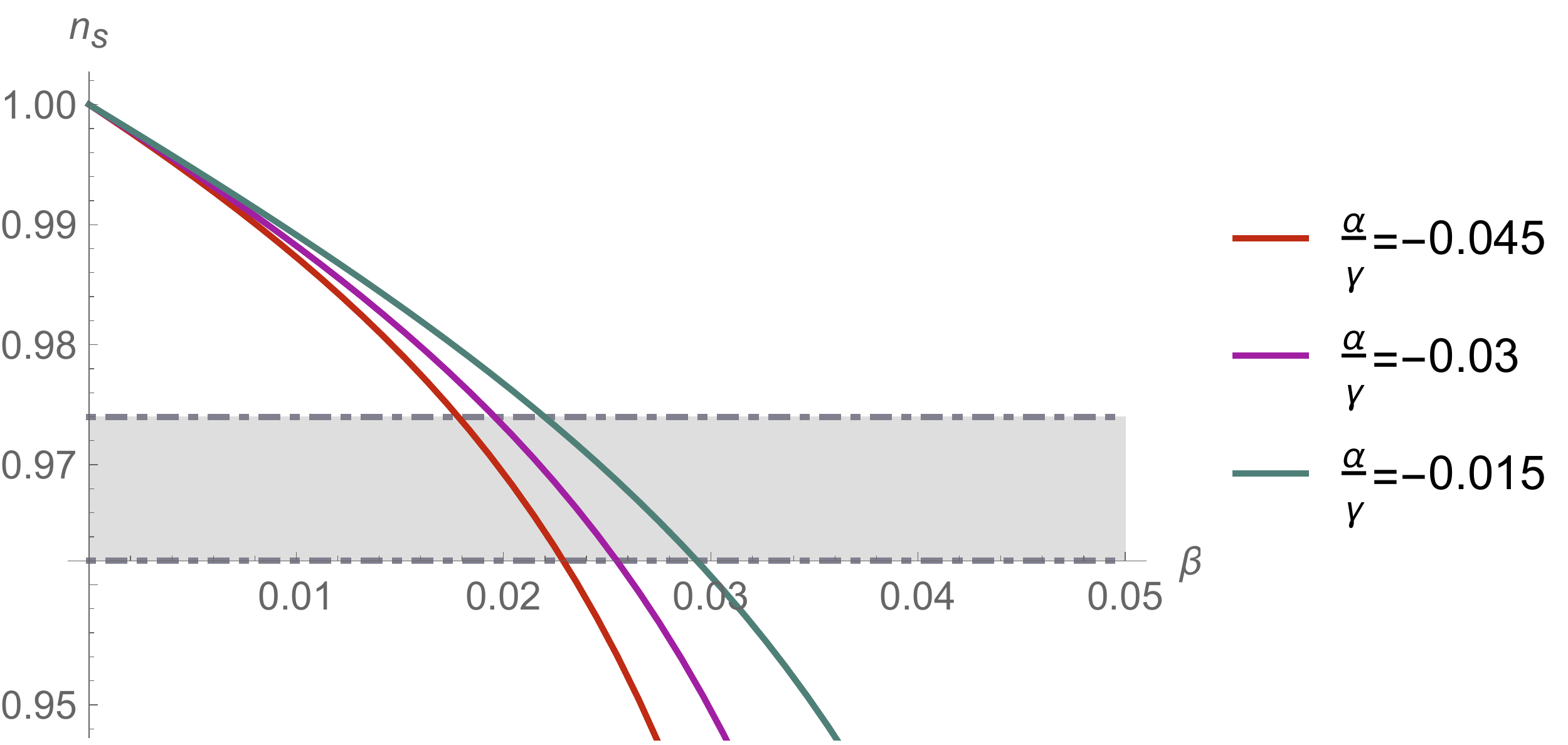}
   \end{minipage}
    \caption{Behavior of the observational indices as a functions of parameter $\beta$ for specific ratios $\Upsilon=\alpha/\gamma$ ($N=60$). Gray regions indicate compatibility of the $n_s$ and $r$ with recent Planck and BICEP2/Keck array data.  }
     \label{fig:1}
\end{figure}

As an example of the inflationary reconstruction, we consider a inflationary model described by the Hubble factor \cite{zhong2018}:
\begin{align}
H(N)=\alpha e^{\beta N}+\gamma.
\label{eq:refname42}
\end{align}
The corresponding mimetic potential and the Lagrange multiplier are given by:

%%%%%%%%%%%
\begin{align}\nonumber
    \lambda(N)&=\frac{1}{   {36 (\gamma +\alpha 
   e^{\beta 
   N})^3}}\Bigg[\alpha  \beta  e^{\beta 
   N} (72
   (\gamma +\alpha 
   e^{\beta 
   N})^4-\frac{d}{{(3 \gamma
   ^2+\alpha ^2 (\beta
   ^2+3 \beta +3) e^{2
   \beta  n}+3 \alpha  (\beta
   +2) \gamma  e^{\beta 
   n})^4}}\\ \nonumber
   & \times\Big[-18 (\beta
   ^2+3 \beta -6) \gamma
   ^6+\alpha ^6 (\beta^2
   +3 \beta +3)^2
   (30 \beta ^2+43 \beta
   +12) e^{6 \beta 
   n}+\alpha ^5 (30 \beta
   ^6+394 \beta ^5+1806 \beta
   ^4+4095 \beta ^3\beta\\ \nonumber 
   & +4941 \beta
   ^2+2961 +648)
   \gamma  e^{5 \beta  n}+3
   \alpha ^4 (3 \beta
   ^6+69 \beta ^5+494 \beta
   ^4+1575 \beta ^3+2502 \beta
   ^2+1920 \beta +540)
   \gamma ^2 e^{4 \beta  n}+3
   \alpha ^3\\  \nonumber&\times (12 \beta
   ^5+142 \beta ^4+725 \beta
   ^3+1698 \beta ^2+1830 \beta
   +720) \gamma ^3 e^{3
   \beta  n}+3 \alpha ^2
   (10 \beta ^4+90 \beta
   ^3+432 \beta ^2+825 \beta
   +540) \gamma ^4 e^{2
   \beta  n}\\ &-9 \alpha  (2
   \beta ^3+3 \beta ^2-37
   \beta -72) \gamma ^5
   e^{\beta 
   n})\Big]\Bigg],
    \label{eq:refname43}
\end{align}
%%%%%%%%%%%
%%%%%%%%%%%
\begin{align}\nonumber
    V(N)&=-\frac{1}{{36
   (\gamma +\alpha 
   e^{\beta 
   N})^3}}\Bigg[-\frac{3 d (\gamma
   +\alpha  e^{\beta 
   N})}{3 \gamma
   ^2+\alpha ^2 (\beta
   ^2+3 \beta +3) e^{2
   \beta  N}+3 \alpha 
   (\beta +2) \gamma  e^{\beta
    N}} -216
   \alpha  \beta  e^{\beta 
   N} (\gamma
   +\alpha  e^{\beta 
   N})^4\\ \nonumber&-\frac{1}{{(3
   \gamma ^2+\alpha ^2
   (\beta ^2+3 \beta
   +3) e^{2 \beta 
   N}+3 \alpha  (\beta
   +2) \gamma  e^{\beta 
   N})^3}}\Big[4 \alpha 
   \beta  d e^{\beta 
   N} (3 \gamma
   ^2+\alpha ^2 (4 \beta
   ^2+9 \beta +3) e^{2
   \beta  N}\\ \nonumber &+\alpha 
   (2 \beta ^2+9 \beta
   +6) \gamma  e^{\beta 
   N}) (3
   (\beta +4) \gamma ^2+4
   \alpha ^2 (\beta ^2+3
   \beta +3) e^{2 \beta 
   N}+\alpha  (2
   \beta ^2+15 \beta
   +24) \gamma  e^{\beta
    N})     \Big] \\ \nonumber &+\frac{d
   (-9 \gamma ^3+3 \alpha
   ^3 (4 \beta ^3+8 \beta
   ^2-3) e^{3 \beta 
   N}+3 \alpha ^2
   (4 \beta ^3+12 \beta
   ^2-9) \gamma  e^{2
   \beta  N}+\alpha 
   (2 \beta ^3+12 \beta
   ^2-27) \gamma ^2
   e^{\beta 
   N})}{(3
   \gamma ^2+\alpha ^2
   (\beta ^2+3 \beta
   +3) e^{2 \beta 
   N}+3 \alpha  (\beta
   +2) \gamma  e^{\beta 
   N})^2}\\ \nonumber &+\frac{1}{(3
   \gamma ^2+\alpha ^2
   (\beta ^2+3 \beta
   +3) e^{2 \beta 
   N}+3 \alpha  (\beta
   +2) \gamma  e^{\beta 
   N})^4}\big(2
   \alpha  \beta ^2 d e^{\beta
    N} (3 \gamma
   +\alpha  (2 \beta +3)
   e^{\beta  N})
   (-9 (\beta +4) \gamma
   ^5\\\nonumber &+28 \alpha ^5 (\beta
   ^2+3 \beta +3)^2 e^{5
   \beta  N}+4 \alpha
   ^4 (7 \beta ^4+72
   \beta ^3+255 \beta ^2+396
   \beta +243) \gamma 
   e^{4 \beta 
   N}\\ \nonumber&+\alpha ^3 (8
   \beta ^4+141 \beta ^3+786
   \beta ^2+1719 \beta
   +1368) \gamma ^2 e^{3
   \beta  N}+3 \alpha
   ^2 (7 \beta ^3+64
   \beta ^2+231 \beta
   +264) \gamma ^3 e^{2
   \beta  N}\\ &+3 \alpha 
   (2 \beta ^2+15 \beta
   +36) \gamma ^4
   e^{\beta 
   N})\big)+72
   \alpha ^2 \beta ^2 e^{2
   \beta  N}
   (\gamma +\alpha 
   e^{\beta 
   N})^4+72
   \alpha  \beta ^2 e^{\beta 
   N} (\gamma
   +\alpha  e^{\beta 
   N})^5\Bigg] .       
   \label{eq:refname44}
\end{align}
%%%%%%%%%%%
Further, the {\it slow-roll} parameters from the perfect fluid approach are:
%%%%%%%%%%%
\begin{align}
    \epsilon=-\frac{\alpha  \beta  e^{\beta  n} ((\beta +6) \gamma +2 \alpha 
   (\beta +3) e^{\beta  n})^2}{4 (\gamma +\alpha  e^{\beta 
   n}) (3 \gamma +\alpha  (\beta +3) e^{\beta  n})^2}
    \label{eq:refname45}
\end{align}
%%%%%%%%%%%
and 
%%%%%%%%%%%
\begin{align}
    \eta=-\frac{\beta  ((\beta +6) \gamma ^2+8 \alpha ^2 (\beta +3) e^{2
   \beta  n}+2 \alpha  (4 \beta +15) \gamma  e^{\beta  n})}{4
   (\gamma +\alpha  e^{\beta  n}) (3 \gamma +\alpha 
   (\beta +3) e^{\beta  n})}.
     \label{eq:refname46}
\end{align}
%%%%%%%%%%%
Finally the corresponding spectral index and the scalar-tensor ratios are listed below:
%%%%%%%%%%%
\begin{align}\nonumber
n_s&=\frac{1}{2
   (\gamma +\alpha  e^{\beta  n}) (3 \gamma +\alpha 
   (\beta +3) e^{\beta  n})^2}  \Big[-3 (\beta ^2+6 \beta -6) \gamma ^3+2 \alpha ^3 (\beta
   +3)^2 (2 \beta +1) e^{3 \beta  n}\\ 
   &+2 \alpha ^2 (2 \beta ^3+16
   \beta ^2+39 \beta +27) \gamma  e^{2 \beta  n}+\alpha  (2
   \beta ^3+3 \beta ^2+12 \beta +54) \gamma ^2 e^{\beta  n}  \Big],
   \label{eq:refname47}
\end{align}
%%%%%%%%%%%
%%%%%%%%%%%
\begin{align}
    r=-\frac{4 \alpha  \beta  e^{\beta  n} ((\beta +6) \gamma +2 \alpha 
   (\beta +3) e^{\beta  n})^2}{(\gamma +\alpha  e^{\beta 
   n}) (3 \gamma +\alpha  (\beta +3) e^{\beta  n})^2}.
    \label{eq:refname48}
\end{align}
%%%%%%%%%%%
The inflationary parameters depend on the e-foldings number $N$ and the three free parameters $\alpha,\beta, \gamma$, which are associated with the Hubble rate (\ref{eq:refname42}). We note that this set can be further simplified by introducing $\Upsilon=\alpha/\gamma$. In Fig. (\ref{fig:1}) we present the behavior of the observational indices as a function of the $\beta$ parameter. For exemplary purposes, we have chosen three values of ratio $\Upsilon$. Gray regions indicate parameter range that is compatible with the BICEP2/Keck array data \cite{ade2016}. Based on the obtained results we observe that the spectral index $n_s$ diminishes with increase of the $\beta$, while scalar to tensor ratio $r$ exhibits increasing behavior. This inflationary model was extensively studied by Zhong and collaborators \cite{zhong2018} in mimetic $f(\mathcal{G})$. We refer reader to their work for the detailed survey of the ranges of parameters that are compatible with the recent observational data provided by BICEP2/Keck array \cite{ade2016}. 

The main goal of this part of our study was to reconstruct proper inflationary model in the mimetic $f(R,R_{\mu\nu}R^{\mu\nu})$ theory. For more detailed discussion of other inflationary scenarios in the mimetic gravity, we refer reader to \cite{zhong2018}. We conclude that extension of the GR presented here is comparable with other mimetic gravities and may consitiute interesting alternative \cite{zhong2018,Odintsov2016}.
%%%%%%%%%%%
%%%%%%%%%%%

%%%%%%%%%%%%%%%%%%%%%%
\section{Summary}
%%%%%%%%%%%%%%%%%%%%%%
In this work, we presented mimetic extension of the $f(R,R_{\mu\nu}R^{\mu\nu})$ gravity using the Lagrange multiplier formalism with the mimetic scalar potential added to the Lagrangian, where mimetic field isolates conformal degree of freedom. Introduced extension was implemented by using the well-known reconstruction technique in order to obtain models of interest. Assuming functional form as $f(R,Y)=R+dY^p$, we obtained the Lagrange multiplier $\lambda$ and the mimetic potential $U(\phi)$ that satisfy the power law evolution of the Universe. Another reconstructed model described symmetric bounce cosmology. Moreover, we have reconstructed model that satisfy transition between matter dominated and accelerated phase in the history of the Universe, showing that this extension of the mimetic gravity is capable to unify various stages of cosmic history. Additionally, using the inverse reconstruction method and choosing appropriate forms of $\lambda$ and $V$ we have obtained $f(R,Y)$ model describing the de-Sitter model with constant Hubble factor $H_0$. Our work is closed with discussion of the inflationary cosmology in a given regime. By using the Hubble parameter $H(N)=\alpha e^{\beta N}+\gamma$ we get Lagrange multiplier and the mimetic potential which successfully describes inflationary model. The reconstructed inflationary model is phenomenologically viable for the wide range of the parameters when confronted with BICEP2/Keck data \cite{ade2016,zhong2018}.

In conclusion, the mimetic extension of the $f(R,T,R_{\mu\nu}T^{\mu\nu})$ can describe any given early or late-time cosmological model in particularly clear way. In general, the procedure presented here can be extended to other higher-curvature modifications of the general relativity or theories that include coupling with the trace of the energy-momentum tensor such as the $f(R,T,R_{\mu\nu}T^{\mu\nu})$ or $f(\mathcal{G},T)$ approaches \cite{sharif2016,zubair2016}. We also remark that the matter fields were not used in our considerations, since main feature of the mimetic gravity is to \textit{mimic} cosmological behavior driven by the matter fields \cite{lambiase2015,nojiri2016}. In the approach presented here, the mimetic condition is supported by the higher curvature terms of the $f(R,R_{\mu\nu} R^{\mu\nu})$ theory, such that the geometry of the considered theory incarnates matter. Moreover, the mimetic condition can play an important role in further considerations of the various higher order extensions of the general relativity and can be extended by inclusion of the unimodular concept in the considerations \cite{nojiri2016}. We also wish to point out that by choosing an appropriate functions and parameters one can obtain analytical results in the mimetic $f(R,R_{\mu\nu}R^{\mu\nu})$ approach, which generally is hard task in the context of the modified theories of gravity \cite{nojrii2017}. Since there are many other proposals of the modified gravities that satisfy realistic models with sufficient degree of accuracy, the local constraints should be employed, such as the Post-Newtonian or Solar-System tests \cite{Vagnozzi2017}. Future studies should be devoted to this topic, by addressing not only mimetic extension presented here, but also other proposals such as mimetic versions of the $f(R,T)$, $f(R)$ or $f(\mathcal{G})$ gravity.
%%%%%%%%%%%%%%%%%%%%%%%%%%%%%%%%%
\appendix
%%%%%%%%%%%%%%%%%%%%%%%%%%%%%%%%%
\setcounter{secnumdepth}{0}
%%%%%%%%%%%%%%%%%%%%%%
\section{Appendix: Full form of the Lagrange multipliers and mimetic potentials}

%%%%%%%%%%%%%%%%%%%%%%
\subsection{The power law evolution}
%%%%%%%%%%%%%%%%%%%%%%
\begin{align}
    \lambda(t)&=\frac{1}{3 n
   (3 n^2-3 n+1)
   t^2}\Bigg[d (-2^{2
   p+1}) 3^p p (8
   p^2-6 p+1) t^2
   (\frac{n^2 (3
   n^2-3
   n+1)}{t^4})^p+d
   n 12^p p (24 p^2-18
   p+1) t^2\\ \nonumber &\times
   (\frac{n^2 (3
   n^2-3
   n+1)}{t^4})^p+6
   n^2 (d 12^p (p-1) p
   t^2 (\frac{n^2 (3
   n^2-3
   n+1)}{t^4})^p-1
   )-18 n^4+18 n^3\Bigg]
\end{align}
%%%%%%%%%%%
%%%%%%%%%%%
\begin{align}
   V(t)=&\frac{1}{(3
   n (3 n^2-3 n+1)
   t^2)}\Bigg[   -d 3^p 4^{p+1} p (8
   p^2-6 p+1) t^2
   (\frac{n^2 (3
   n^2-3
   n+1)}{t^4})^p\\ \nonumber&+3
   n^2 (d 12^p (8
   p^2-8 p+3) t^2
   (\frac{n^2 (3
   n^2-3
   n+1)}{t^4})^p-8
   )\\ \nonumber&-d n 3^{p+1} 4^p
   (16 p^3-4 p^2-4
   p+1) t^2
   (\frac{n^2 (3
   n^2-3
   n+1)}{t^4})^p +162 n^5-234 n^4\\ \nonumber&+9
   n^3 (d 12^p (p-1) t^2
   (\frac{n^2 (3
   n^2-3
   n+1)}{t^4})^p+1
   4)\Bigg]
\end{align}
%%%%%%%%%%%
%%%%%%%%%%%%%%%%%%%%%%%%%%%%%%%%%
\subsection{The symmetric bounce}
%%%%%%%%%%%%%%%%%%%%%%%%%%%%%%%%%
%%%%%%%%%%%
\begin{align}
&\lambda(t)=2 \Big[2 \alpha +8 \alpha ^2 d p (48 \alpha ^2+576 \alpha ^4 t^4+288 \alpha ^3 t^2)^{p-1}\frac{-6 \alpha  d (p-1) p t^2 (\alpha  t^2-2)
   (4 \alpha  t^2+1) (48 \alpha ^2+576 \alpha ^4 t^4+288 \alpha ^3 t^2)^p}{(12 \alpha ^2 t^4+6 \alpha  t^2+1)^2}\\ \nonumber 
   &+\frac{d 48^p (p-1)
   p (3 \alpha  t^2+1) (\alpha ^2 (12 \alpha ^2 t^4+6 \alpha  t^2+1))^p (48 \alpha ^3 (4 p-5) t^6+12 \alpha ^2 (8 p-9) t^4+6
   \alpha  (2 p-1) t^2+1)}{(12 \alpha ^2 t^4+6 \alpha  t^2+1)^3})\Big],
\end{align}
%%%%%%%%%%%
%%%%%%%%%%%
\begin{align}\nonumber
    V(t)&=d 48^p (\alpha ^2 (12 \alpha ^2 t^4+6 \alpha  t^2+1))^p+\alpha ^5 d 2^{4 p+3} 3^{p+1} (p-1) p t^2
   (2 \alpha  t^2+3) (4 \alpha  t^2+1) (\alpha ^2 (12 \alpha ^2 t^4+6 \alpha 
   t^2+1))^{p-2}\\ 
   &-576 \alpha ^4 d p t^4 (48 \alpha ^2+576 \alpha ^4 t^4+288 \alpha ^3
   t^2)^{p-1}   -12 \alpha  (4 \alpha  t^2+1)-4 \alpha 
   (6 \alpha  t^2+1)\\\nonumber &+\frac{d 3^p 4^{2 p+1} (p-1) p (3 \alpha  t^2+1) (48 \alpha ^3 (4 p-5) t^6+12 \alpha ^2
   (8 p-9) t^4+6 \alpha  (2 p-1) t^2+1) (\alpha ^2 (12 \alpha ^2 t^4+6 \alpha 
   t^2+1))^p}{(12 \alpha ^2 t^4+6 \alpha  t^2+1)^3}.
\end{align}
%%%%%%%%%%%

\subsection{Transition between matter-dominated and accelerated phases}
For the small $t$ limit ($g_0 \rightarrow 0$):
%%%%%%%%%%%
\begin{align}
    \lambda (t)&=\frac{1}{3 g_1 \left(3
   g_1{}^2-3 g_1+1\right) t^2}\Bigg[d \left(-2^{2
   p+1}\right) 3^p p \left(8
   p^2-6 p+1\right) t^2
   \left(\frac{g_1{}^2 \left(3
   g_1{}^2-3
   g_1+1\right)}{t^4}\right){}
   ^p+d g_1 12^p p \left(24
   p^2-18 p+1\right)\\& t^2
   \left(\frac{g_1{}^2 \left(3
   g_1{}^2-3
   g_1+1\right)}{t^4}\right){}
   ^p+6 g_1{}^2 \left(d 12^p
   (p-1) p t^2
   \left(\frac{g_1{}^2 \left(3
   g_1{}^2-3
   g_1+1\right)}{t^4}\right){}
   ^p-1\right)-18 g_1{}^4+18
   g_1{}^3\Bigg],
\end{align}
%%%%%%%%%%%
%%%%%%%%%%%
\begin{align}\nonumber
   V(t)&= -\frac{1}{3 g_1 (3
   g_1{}^2-3 g_1+1) t^2}\Bigg[d 3^p 4^{p+1} p (8
   p^2-6 p+1) t^2
   \Big(\frac{g_1{}^2 (3
   g_1{}^2-3
   g_1+1)}{t^4}\Big){}
   ^p+3 g_1{}^2 \Big(d 12^p
   (8 p^2-8 p+3)
   t^2
   \\ \nonumber &\times\Big(\frac{g_1{}^2
   (3 g_1{}^2-3
   g_1+1)}{t^4}\Big){}
   ^p-2\Big)-d g_1 3^{p+1}
   4^p (16 p^3-4 p^2-4
   p+1) t^2
   \left(\frac{g_1{}^2 (3
   g_1{}^2-3
   g_1+1)}{t^4}\right){}
   ^p\\&+\Big(9 g_1{}^3 (d 12^p
   (p-1) t^2
   \Big(\frac{g_1{}^2 (3
   g_1{}^2-3
   g_1+1)}{t^4}\Big){}
   ^p+4\Big)+54 g_1{}^5-72
   g_1{}^4\Bigg]
\end{align}
%%%%%%%%%%%

In the large $t$ limit ($g_1 \rightarrow 0$):
%%%%%%%%%%%
\begin{align}
    \lambda(t)=0,\;\;\; v(t)=6 g_0{}^2-d 36^p (p-1)
   \left(g_0{}^4\right){}^p.
\end{align}
%%%%%%%%%%%

\bibliographystyle{apsrev}
\bibliography{bibliography}

%%%%%%%%%%%%%%%%%%%%%%%%%%%%%%%%%%%%%%%%%%%%%%%%%%%%%%%%%%%%%%%%%%%%%%%%%%%%%%%%%%%%%%%%%%%%%%%%%%%%%%%

%%%%%%%%%%%%%%%%%%%%%%%%%%%%%%%%%%%%%%%%%%%%%%%%%%%%%%%%%%%%%%%%%%%%%%%%%%%%%%%%%%%%%%%%%%%%%%%%%%%%%%%%%%%%%%%%%%%%%%%%%%%%%%%%%%%%%%%%%%%%%%%%%%%%%%%%%%%%%%%%%%%%%%%%

\end{document}